\documentstyle[sprocl2,epsfig]{article}
\def\be{\begin{equation}}
\def\ee{\end{equation}}
\def\bea{\begin{eqnarray}}
\def\eea{\end{eqnarray}}

\begin{document}
\title{Spatial 't Hooft loop to cubic order in hot QCD II}
\author{P. Giovannangeli}
\address{Department of Theoretical Physics, University of Bielefeld,
Germany\\
E-mail: giovanna@physik.uni-bielefeld.de} 

\author{C. P. Korthals Altes}

\address{Centre Physique Th\'eorique au CNRS, Case 907, Luminy,F13288, Marseille, France\\ E-mail:altes@cpt.univ-mrs.fr}  

\maketitle

\abstracts{In this paper we provide an exact formula for the area law of the 't Hooft loop for any
SU(N) gauge group to cubic order in hot gauge theory.
 The correction is very small for all temperatures above $T_c$, in stark contrast to the cubic correction to the pressure. 
The gradient approximation in a previous paper, only valid for large N, is in excellent agreement with the present, exact evalution. Comparison to lattice data is good. Casimir scaling is violated by a small amount not yet resolved by the precision of lattice data.}

\section{Introduction}

In the study of hot QCD there is a natural quantity, the electric flux loop or 't Hooft
loop,  that measures the electric flux in the plasma. Its behaviour is governed by an area law in the hot phase, and, if quark fields are decoupled, a perimeter law in the cold phase. Lattice simulations~\cite{deforluc} start to determine with considerable accuracy this quantity.
 
The coefficient in the area law, the ``tension'', is calculable in perturbation theory, in the same sense as the pressure~\cite{braaten}. 

Unlike for the pressure the loop is intimately related to a global,
discrete symmetry, $Z(N)$ symmetry, whose order parameter is the 
thermal Polyakov loop, winding around the 4d system in the periodic temporal or ``thermal'' direction. 

Like for the pressure there is a  loop expansion. But the propagators and vertices in the loops are now propagating in the background of the Polyakov loop.
Thus we derive an effective potential in terms of the Polyakov loop.

Z(N) symmetry tells us the effective action is the same for all values of the
Polyakov loop that differ by a Z(N) phase. In particular the minima of the potential are determined by the value of the loop being a Z(N) phase. The minima
of the effective potential
have the value of the pressure. In between the minima the system can tunnel,
and this gives rise to the area law for the flux loop.

The pressure has a one loop contribution, the Stefan-Boltzmann result. So has the flux loop. The first correction is the two loop result. This is known since long~\cite{bhatta}. The cubic result has been derived in the same spirit
as for the pressure. The cubic result in the pressure reflects the appearance of the Debye mass. In the effective potential the Debye mass acquires dependence on the Polyakov loop and this gives rise to a natural limitation. The limitation is due to the effective potential being concave as function of the Polyakov loop, far enough from the minima.
This means its second derivative is negative, giving an unstable propagator
with a negative mass-squared. Nethertheless for a large number of colours one can still establish a result for the cubic correction to the flux loop. This was done in the preceding paper~\cite{i}, henceforth referred to as I.  This result was in agreement with 
 a recent numerical determination~\cite{deforluc}  of the flux loop average
in SU(4).

In this sequel  we give explicit
results for any number of colours. The method we use is is explained in section~\ref{sec:frame}. More details on the method follow in section~\ref{sec:uv}.
Predictions are reported in section~\ref{sec:predict}, and are compared 
with our previous work for large number of colours in the next section. The last section concludes.

\section{Framework}\label{sec:frame}

The starting point of our investigation is the electric flux loop, as defined in I for $SU(N)$ gauge theory:
\be
V_k(L)=\exp{i{4\pi}\int_{S(L)} dxdyTrE_zY_k}.
\label{eq:loop}
\ee

The loop $L$ is taken in the x-y plane, spanning a minimal surface $S(L)$.
The electric field strength $E_k$ and gauge potential $A_k$ are written in $N\times N$ matrix form
 $E_k=E^a_k\lambda_a$. Quantization is obtained through $[E_k^a(\vec x),A_l^b(\vec y)]={1\over i}\delta^{a,b}\delta_{k,l}\delta(\vec x-\vec y)$\footnote{ Normalization
of the Gell-Mann matrices is as usual:
$Tr \lambda_a\lambda_b=2\delta^{a,b}$
~~\mbox{and}~~ $[\lambda_a,\lambda_b]=if_{abc}\lambda_c$, $f_{abc}f_{a'bc}= \delta^{a,a'}$.}. 

The $N\times N$ matrix $Y_k$ is a generalization of the hypercharge matrix. It generates elements of the Z(N) centergroup of SU(N):
\be
\exp{i2\pi Y_k}=\exp{-i{2\pi\over N} k}{\bf 1}
\label{center}
\ee
and can be taken explicitely as:
\be
Y_k=\mbox{diag}(N-k,N-k,....N-k,-k,-k,...-k).
\ee
To be traceless $Y_k$ has k entries $N-k$ and $N-k$ entries $-k$.

Obviously, $V_k$ is an electric flux loop operator and unitary. Because of the 
canonical commutation relations $V_k$ transforms a spatial Wilson loop in the
fundamental representation by the centergroup factor eq. (\ref{center}). 
This property ( the 't Hooft commutation relation~\cite{thooft78}) does not depend on the surface $S(L)$.

Indeed one can show that $V_k$  is  a gauge transformation, with a discontinuity $\exp{i{2\pi\over N}}$ when going clockwise around the loop $L$~\cite{korthalskovnersteph}, i.e. the 't Hooft loop~\cite{thooft78} which is the operator creating a magnetic Dirac vortex loop  of strength $\exp{i{2\pi\over N}}$. 

The thermal average of the loop is the Gibbs trace over physical states:
\be
\langle V_k(L)\rangle=Tr_{phys}V_k(L)\exp{(-{H\over T})}/Tr_{phys}\exp{(-{H\over T})}.
\label{average}
\ee

In the hot phase the average is an area law:
\be
\langle V_k(L)\rangle=\exp{-\rho_k(T)A(L)}.
\label{arealaw}
\ee

The coefficient $\rho_k(T)$ is the 't Hooft k-tension.
Our interest is mainly in the dependence on the strength $k$.

\subsection{Pathintegral for the thermal average}

This average can be reexpressed along familiar lines in terms of  a path integral. 

As a surface of electric dipoles in the x-y plane our loop will cause, like in static classical electrodynamics, a profile for $A_0(z)$ that has a discontinuity across the dipole layer, and dies off at infinity. Its slope is continuous
everywhere, including at the layer.  

In the non-Abelian case one must find a gauge invariant version of 
$A_0(\vec x )$.
This is the set  of eigenvalues of the time ordered Polyakov line:
\be
P(A_0(\vec x))={1\over N}{\cal{P}}\exp{i\int_0^{1/T} d\tau A_0(\tau,\vec x)}.
\label{polline}
\ee
This is because under a periodic gauge transformation $\Omega$:
\be
P(A_0^{\Omega})=\Omega P(A_0)\Omega^{\dagger}.
\label{invpolline}
\ee

Under a gauge transformation  $\Omega_l$ with a discontinuity $\exp{il{2\pi\over N}}$ in the periodic time direction the eigenvalues will be shifted over Z(N)  angles. So in the high T, broken Z(N) symmetry phase the Polyakov line is an orderparameter. The eigenvalues will be written as $2\pi q_i$, with $i=1,..,N$ and $\sum_iq_i=0$.

The average of the loop will then be given by a path integral over all possible
profiles:
\be
\langle V_k(L)\rangle=\int DQ(z)\int DA \delta(\exp{(i2\pi Q (z))}-\overline P(A_0))\exp{-{1\over {g^2}}S(A)}
\ee
where $Q$ is the diagonal NXN matrix with elements $q_i$. The Polyakov line $\overline P(A_0)$ is averaged over the transverse directions $x$ and $y$, as indicated by the bar.

Doing the path integral over the vector potentials gives the effective action:
\be
\int DA \delta(\exp{(i2\pi Q(z))}-\overline P(A_0))\exp{-{1\over {g^2}}S(A)}=exp{-L_{tr}^2S_{eff}(q)}.
\label{effaction}
\ee
The effective action has the Z(N) shift symmetries mentioned below eq.(\ref{invpolline}).

The transverse size of the system is $L_{tr}$. If the loop is very large, comparable to the transverse size we can equate $A(L)=L_{tr}^2$ in eq. (\ref{arealaw}). 

 As we are interested in the thermodynamic limit $L_{tr}\rightarrow \infty$ the integral over the profile $Q$
reduces to minimizing the effective action over the profile with the boundary
conditions $Q=0$ at $z=\pm\infty$. The symmetries of  the effective action
ensure that the minimizing path from $\exp{i2\pi Q}=1$ to $\exp{i{2\pi\over N}k}$ is the straight path from $Q=0$ to $Q=Y_k$ parametrized by $qY_k$, $q$ ranging from 0 to 1.

This simplifies the task enormously. In terms of the single parameter $q$
we have to minimize the effective action with boundary condition $q=0$
at $z=\pm\infty$ and $q$ making a jump of ${2\pi\over N}k$ at the surface of the loop.

Finally we find the 't Hooft tension in terms of the effective action by combining eq. (\ref{arealaw}), eq. (\ref{effaction}) and the remark underneath the latter to write:
\be
\rho_k(T)=min_k S_{eff}(q)
\label{tensioneff}
\ee
where the r.h.s. is understood to be minimized on the path $qY_k$.

As explained in the following sections, one can evaluate the tension at high temperature in a perturbative expansion, with odd powers of $g$:
\be
\rho_k=\rho_k^{(1)}+\rho^{(2)}_k+\rho^{(3)}_k+....
\label{expansion}
\ee

The purpose of this paper is to compute $\rho^{(3)}_k$.

\subsection{The effective action in perturbation theory}

The effective action can be computed in a loop expansion, in precise analogy with the perturbative calculation of the pressure. The  diagrams
are  q-dependent through the propagators and vertices, due to the delta function constraint in  eq. (\ref{effaction}).

Without fluctuations the effective action will just equal 
\be
S_{eff}(q)={1\over{g^2}}T\int dzTr(2\pi\partial_z q Y_k)^2
\ee
in terms of the classical electric field strength $E_z=\partial_zA_0$.
 Minimize this with the system having extension L in the z-direction. Keeping the kinetic term continuous, whereas $q$ jumps
, gives a contribution where $q(z)=az+b$ with $a=O({1\over L})$. Hence the integral over $z$ will render  $S_{eff}(q)=O({1\over L})$.

So we need quantum fluctuations for the tension. Taking those into account introduces the 
screening length $m_D^2={g^2NT^2\over 3}$ for $A_0$ (or the Polyakov line).
This means that $A_0$ and therefore $q$ will vary only appreciably over distances $l_D=m_D^{-1}$.
To one loop order the effective action becomes the logarithm of the determinant of the fluctuation matrix in the action $S(A)$ in eq. (\ref{effaction}):
\be
\det(-D_{\mu}^2(q)).
\label{propagator}
\ee

We suppressed  colour indices and gauge parameter dependence, which does not survive the determinant anyway. We have  $D_{\mu}(q)=\partial_{\mu}-i\delta_{\mu,0}[Q_{\mu},$ and $Q_0=2\pi Tq Y_k$ from expanding the action $S(A)$ around $Q$
to quadratic order. 

Note that the propagators ${-1\over {D^2(q)}}$ have a background dependence.
In the Cartan basis~\footnote{Given by the set of N(N-1) generators $\lambda_{ij}$ with only one non-zero element ${1\over{\sqrt{2}}}$ in the (ij) entry, and N-1 diagonal $\lambda_d=(\sqrt{2d(d-1)})^{-1/2}\mbox{diag}(1,1,....,1,1-d,0,...0)$} propagators with index $i\neq j$ have a mass $q^2$ if
$(Y_k)_{ii}-(Y_k)_{jj}\neq 0$, because of the commutator in the covariant derivative. In all other cases the induced background mass is zero.

In computing the determinant we can neglect the slow variation of the backgound and get~\cite{bhatta}:
\be 
S_{eff}={1\over{g^2}}T\int dz Tr(2\pi\partial_z q Y_k)^2+V_1(q)
\label{effaction1loop}
\ee
with
\be
V_1(q)={4\pi^2T^3\over 3}k(N-k)\int dz q^2\big(1-|q|\big)^2.
\label{onelooppot}
\ee

This potential is periodic in $q$ mod 1 as required by the Z(N) symmetry
 discussed in the previous subsection.

Upon minimization one gets from eq. (\ref{tensioneff}, (\ref{effaction1loop})  and (\ref{onelooppot}):
\be
\rho_k(T)={4\pi^2\over{3\sqrt{3g^2N}}}k(N-k)T^2 
\label{oneloopresult}
\ee
as one loop result, with Casimir scaling in the strength of the loop.

The one loop profile reads:
\be
q_m(z')={\exp{z'}\over{1+\exp{z'}}}
\label{profile}
\ee
with $z'=m_Dz$.

We have, for notational convenience, lifted the branch of the profile on the right of the loop by one unit, rendering it continous and ending up at $q(z=\infty)=1$, not anymore at $q(z=\infty)=0$ . It is for this  reason that the tension is often referred to as a domainwall tension between two regions of space
with different values for the Polyakov line, viz. $1$, and $\exp{ik{2\pi\over N}}$. This is a tenable interpretation of the Euclidean lattice simulation.
In a realistic plasma however the physical meaning of the Polyakov line is not clear.

 This change in the profile does not alter the minimization procedure because the kinetic energy is continuous, and $V_1(q)$ has the required periodicity. The tension does not change.

The two loop contribution will add $g^2V_2(q)$ to the effective action. It will also renormalize the kinetic term in the effective action and introduce renormalization effects in the coupling~\cite{bhatta}. Quite miraculously Casimir scaling survives in two loops~\cite{giovanna}.

\subsection{ $g^3$ contribution to 't Hooft tension}\label{subsec:cubiccont}

\begin{figure}
\begin{center}
\includegraphics{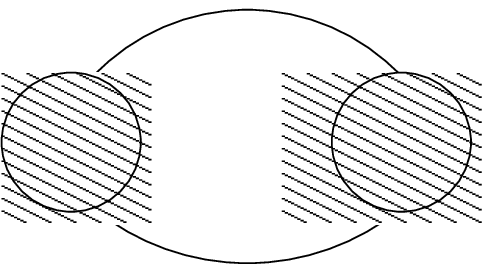}
\caption{The only  three loop diagram of  free energy topology with an infrared divergence. The shaded blob is the one loop selfenergy. The colour index of the two propagators need not be the same due to background dependence inside the blobs.}
\label{fig:Df}
\end{center}
\end{figure}

Formally the three loop free energy graphs  are of $O(g^4)$. But due to infrared divergencies a certain subset of three loop diagrams is only $O(g^3)$ and this order  the only one of interest here. They are shown in figure (\ref{fig:Df}).

\begin{figure}
\begin{center}
\includegraphics{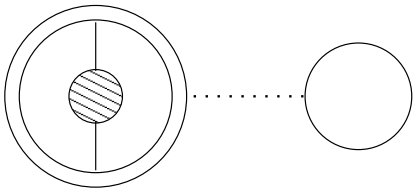}
\caption{Only three loop insertion diagram with infrared divergence}
\label{fig:Dp}
\end{center}
\end{figure}
 In section 5.1 of I we analyzed the order $g^3$ contributions. They originate 
in the propagators sticking into the blobs. If they have no background induced mass then there is a linear divergence.

 Note that the same happens for the pressure for {\it all} ~lines.  Because the pressure is evaluated at $q=0$,  no line has a background induced mass. All the zero-momentum blobs in this case equal the Debye mass $m_D^2$. And when we integrate the spatial momenta in the graph one finds $L^3(N^2-1)\int d\vec p{1\over{(\vec p)^2}} m_D^4=m_D^3$. The point is that {\it all}~ graphs with an arbitrary number of selfenergy insertions are of the same order, so have to be resummed.

This leads for the free energy  to a cubic contribution:
\be
V{f_3\over T}={1\over 2}(N^2-1)\log\det(-\vec\partial^2+m_D^2)
\label{pressure}
\ee

Returning to the q-dependent case,
the value of the selfenergy blob where a line with colour $a$ enters, and a line with colour $b$ leaves (both with zero momentum) is denoted by $\Pi(k,q)^{a,b}$, the self-energy matrix. In I we showed  this is gauge choice independent. In the limit of $q=0$ it reduces obviously to $m_D\delta_{a,b}$. The indices a and b run, as said before, through those indices that have no background mass induced in the corresponding propagator. In the 
channel $qY_k$ this corresponds to a total number of such indices $N_0(k)=N^2-1-2k(N-k)$. This is the dimension of the self energy matrix.

The matrix $\Pi(k,q)_{a,b}$ can be diagonalized. This is done in I, Appendix C.
The result is, using the notation  $r={k\over N}$:
\begin{itemize}
\item{ (k-1)(k+1) eigenvalues equal $m_D^2(1+6(1-r)q(q-1))$}
\item{(N-k-1)(N-k+1) eigenvalues equal $m_D^2(1+6rq(q-1))$}
\item{one eigenvalue equals $m_D^2(1+6q(q-1))$.}
\end{itemize}
The total number of eigenvalues corresponds indeed to $N_0(k)$.

In the presence of the background $q$ the free energy $V{f_3}$ changes into:
\bea
V{f_3}+F^{(k)}_3(q)&=&{1\over 2}(N^2-1)\log \det(-\vec\partial^2+m_D^2)\\ \nonumber
&+&{1\over 2}\Big(
\log\det(-\vec\partial^2\delta_{a,b}+\Pi(k,q)_{a,b})-
\log\det(-\vec\partial^2\delta_{a,b}+\Pi(k,q=0)_{a,b})\Big).
\label{cubicdets}
\eea
The last term is subtracted to avoid double counting in the first term.
We used a dimensional argument that the first term  on the r.h.s. of 
eq. (\ref{cubicdets}), the cubic order of the pressure, is  proportional to $(N^2-1)L^3m_D^3$. 

The second and third term both contain the same $N_0(k) L^3m_D^3$ contribution, which cancels in the difference. This contribution will show up later as the leading contribution in the density of states (see the discussion below eq. (\ref{diffeqn})). But what remains in the difference is $O(L^2m_D^2)$ i.e. a surface contribution. It is this surface contribution that we will calculate.

Thus we can put $q=q_m$, because corrections to $q_m$ will give higher order corrections to the surface term in (\ref{cubicdets}).

The cubic contribution  $\rho_k^{(3)}$ to the tension is, abbreviating $m_D^2(q_m)= m_D^2(1+6rq_m(q_m-1))$:

\bea
\nonumber
 \rho_k^{(3)}L^2&=&{1\over 2}(k^2-1)\Bigg(\log\det\Big(-\vec\partial^2+m_D^2(q_m)\Big)-{1\over 2}
\log\det(-\vec\partial^2+m_D^2)\Bigg) \\
\nonumber  &+&k\leftrightarrow N-k \\ 
 &+&{1\over 2}\log\det\Big(-\vec\partial^2+m_D^2(q_m)\Big)-{1\over 2}\log\det\Big(-\vec\partial^2+m_D^2\Big).
\label{eq:logdetcu}
\eea

A few comments are in order to transform eq.(\ref{eq:logdetcu}). We work with dimensional regularization, with $d\rightarrow 3$.

\begin{itemize}
\item{Use $\log\det=Tr\log$ and $ Tr\log X-Tr\log Y=\int_0^{\infty}{dt\over t}(exp{-tY}-\exp{-tX})$}
\item{Use the Poisson formula $\sum_n\exp{-t({2\pi n\over{L_{tr}}})^2}={1\over{\sqrt{4\pi t}}}\sum_l\exp{-{(lL_{tr})^2\over {4t}}}$ for the periodic transverse directions.}
\end{itemize}

A typical term in   eq. (\ref{eq:logdetcu}) reads then:
\bea
\nonumber g(r,d)&=&-{1\over 2}\log\det(-\vec\partial^2+m_D^2(q)-{1\over 2}
\log\det(-\vec\partial^2+m_D^2)\\ 
\nonumber &=&{1\over 2}L^{(d-1)}\int_0^{\infty}{dt\over t}(4\pi t)^{{-d+1\over 2}}\Big(\sum_n\exp{(-n^2L^2/4t)}\Big)^{(d-1)}\Big\{Tr \exp{-t\big(-\partial_z^2+ m_D^2(q)\big)}\\ &-{1\over 2}&Tr \exp{-t\big(-\partial_z^2+ m_D^2\big)}\Big\}.
\label{trlog}
\eea
The first trace on the r.h.s.  necessitates the evaluation of the eigenvalue spectrum of a one dimensional Schroedinger operator:
\be
-\partial_z^2+ m_D^2(1+6rq_m(q_m-1)).
\label{schroe}
\ee

\section{Solving the Schroedinger equation}\label{sec:uv}

The Schroedinger equation has as potential term
\be
m_D^2(1+6rq_m(q_m-1))={m_D^2\over 4}(4-6rsech^2({z\over 2}))
\ee

where we used eq. (\ref{profile}). This is a solvable potential~\cite{morseref}.
With $6r=j(j+1)$ we have:
\begin{equation}
[-\partial_z^2+{m_D^2\over 4}(4-j(j+1)sech^2({z\over 2}))]\psi(z)=E(j)\psi(z).
\label{diffeqn}
\end{equation}

For any integer $j$ we have a bound state entering with parity $(-)^j$ and binding energy $E^{\pm}_b(j)$, the superscript referring to parity. There is a continuous spectrum, parametrized by $E=p^2+m_D^2$. The momentum $p$ runs from $0$ to $\infty$.

The density of states in the positive (negative) parity channel is $\nu^+-\nu_0^+$ ($\nu^--\nu^-_0$). The superscript $0$ refers to the density of states in the absence of the potential (proportional to ${L\over {\pi}}$), and should be subtracted according to eq.(\ref{trlog}).

    These densities  of states can be expressed in terms of the respective
phase shifts $\cos(pz+\delta^+)$ ($\sin(pz+\delta^-)$). In terms of a dimensionless momentum $k$ defined by  $pz=k {zm_D\over 2}$ the phase shifts are related to the densities by~\cite{barton}\footnote{We are indebted to R.L. Jaffe for this reference}:
\bea
\int^{\infty}_0{dk\over{\pi}}(\nu^+(j,k)-\nu_0^+)&=&\int^{\infty}_0{dk\over{\pi}}({\partial\delta^+(j,k)\over{\partial k}}-{1\over 2}(1-\delta_{[j]+1,j})\delta(k)\\
\int^{\infty}_0{dk\over{\pi}}(\nu^-(j,k)-\nu_0^-)&=&\int^{\infty}_0{dk\over{\pi}}{\partial\delta^-(j,k)\over{\partial k}}.
\label{densityphase}
\eea
The integer part of $j$ is written as $[j]$, and $[j]<j\le [j]+1$. 
Note the  delta function in the positive parity channel. It is only  there when $j\neq ~\mbox{integer}$~\cite{barton}.

The positive parity bound states are for fixed $j$:
\be
 E_b^+(j)={m_D^2\over 4}(4-(j-n)^2) 
\label{eplus}
\ee
\noindent with n a non-negative even integer, and $n\le [j]$.
Their negative parity partners are ($n$ odd positve integer) 
\be
 E_b^-(j)={m_D^2\over 4}(4-(j-n)^2).
\label{eminus}
\ee

This is important for the correct ultraviolet behaviour of the integral (\ref{trlog}) as we will see in the next subsection.

The phase shifts are given explicitely in  appendix A.

We combine the knowledge of the eigenvalues and the relation of density of states to phase shifts in a formula for the cubic tension, by plugging eq. (\ref{densityphase}) into eq. (\ref{trlog}):
\begin{eqnarray}
\nonumber g(j,d) & = &-{1\over 2} (Lm_D/4)^{(d-1)}\int{dv\over v}(\pi v)^{{-d+1\over 2}}\big(\sum_n\exp{(-n^2(Lm_D/4)^2/v)}\big)^{(d-1)}\\ 
\nonumber       & \times &\big(\exp{-\widetilde E^+_b(j)v}+\Theta(j-1)\exp{-\widetilde E_b^-(j)v}\\ 
       & + & \int_0^{\infty} {dk\over{\pi}}\big({\partial\delta(j,k)\over{\partial k}}-(1-\delta_{[j]+1,j}){\pi\over 2}\delta(k)\big)\exp{-(k^2+4)v}\big). 
\label{typical}
\end{eqnarray}

The $\Theta(x)=0$ for $x\le 0$, and $\Theta(x)=1$ for $x>0$.
The eigentime $v=tm_D^2/4$, and the bound state energies $\tilde E$ are rescaled by the same factor $m_D^2/4$ in (\ref{eplus}) and (\ref{eminus}). 

The phase shift $\delta(j,k)=\delta^+(j,k)+\delta^-(j,k)$ and  we have absorbed a factor $m_D^2/4$ into the proper time $t$
and called the new variable $v$.

\subsection{Ultraviolet behaviour}

The ultra-violet behaviour of the tension cannot be worse than that of the pressure in eq. (\ref{cubicdets}). And once we have subtracted the ultraviolet divergencies from the temperature induced phenomena only infrared infinities may survive.

 The pressure in dimensional regularisation
has no ultraviolet infinity. This is best seen by writing the pressure in terms of the proper time t:
\be
p_3={1\over 2}\int_0^{\infty}{dt\over t}(4\pi t)^{{-d+1\over 2}}\exp{-tm_D^2}.
\ee
At t=0 the integrand behaves like $t^{-1/2}$ and produces a $\Gamma(-{3\over 2})$. Thus dimensional regularization subtracts in odd dimensions the u.v. infinities, i.e. produces no poles.

Let us now look at the expression for a typical contribution to  $\rho_k^{(3)}$, eq. (\ref{typical}). We get for the integrand at $v=tm_D^2/4=0$, with $n_b$ the number of bound states:
\be
\Big(n_b-({\delta(j,k=0)\over{\pi}}+(1-\delta_{[j]+1,j}){1\over 2})\Big)v^{{d-3\over 2}}.
\label{levinson}
\ee

This would give for d=3 an ultraviolet divergence. There is, however, a connection between the number of bound states and the phase shifts at $k=0$, Levinson's theorem~\cite{barton}. It tells us that the coefficient is zero for all j!

With the explicit formulae of the previous section we have
  $\delta^+(j,k=0)=[j]{\pi\over 2}$, with $[j]=(2n+1)~~\mbox{if}~~j>2n$, n integer. In particular, in between j=0 and j=2, the number of positive parity bound states is 1, and ${\delta^+(j,0)\over{\pi}}+{1\over 2}=1$. 

For the negative parity phase shifts we have ${\delta^-(j,0)\over{\pi}}=\Theta(j-1)=n_b^-$ for $j\le 3$. 

More generally the integrand in eq.(\ref{typical}) admits an expansion for small $v$ in powers of $v^{n/2}$, n a positive integer. However for even n the powers in v from the bound states cancel with those from the phase shift integral,
and only half integral powers stay. This is non-trivial for the case of j non-integer, the main reason being that there is not only transmission but also reflection. The reader can find the relevant formulae in  the appendices.
For the convenience of the reader we give the formula that results  after the cancellation of the integer powers: 

\begin{eqnarray}
\nonumber & &\big(\exp{-\widetilde E^+_b(j)v}+\Theta(j-1)\exp{-\widetilde E_b^-(j)v}\\ 
\nonumber& + & \int_0^{\infty} {dk\over{\pi}}\big({\partial\delta(j,k)\over{\partial k}}-(1-\delta_{[j]+1,j}){\pi\over 2}\delta(k)\big)\exp{-(k^2+4)v}\big)\\ 
&=&{2\over{\sqrt{\pi}}}\int_0^{\infty}du\exp{-u^2}\sinh((j+1)u\sqrt{v}){\sinh(ju\sqrt{v})\over{\sinh(u\sqrt{v})}}+I_0(j,v).
\label{general}
\end{eqnarray}

The remainder  $I_0(j,v)$ is identically zero, as shown in  appendix D.

The conclusion is that ultraviolet divergencies are absent for any value of $j$.

For the special case of j integer there is only reflection. 
Then,  as
 noticed by Muenster~\cite{muenster}, one can  express the integrand in terms of the error function:
\begin{equation}
Erf(\sqrt{v})={2\over{\sqrt{\pi}}}\int_0^{\sqrt{v}}dx\exp{-x^2}.
\label{errorfunction}
\end{equation}

His relation is, for j=1:

\begin{equation}
\exp{(-vE_b^+(j=1))}+\int_0^{\infty} {dk\over{\pi}}{\partial\delta(j=1,k)\over{\partial k}}\exp{-(k^2+4)v}=\exp{(-3v)}Erf(\sqrt{v}).
\label{mue1}
\end{equation}
The r.h.s. has an expansion in $(\sqrt{v})^{2n+1}$, n=0,1,2,...

 Analogously, for j=2 we have :

\bea
\nonumber \exp{(-vE_b^+(j=2))}&+&\exp{(-vE_b^-(j=2))}+\int_0^{\infty} {dk\over{\pi}}{\partial\delta(j=2,k)\over{\partial k}}\exp{-(k^2+4)v}\\ 
&=&Erf(2(\sqrt{v})+\exp{(-3v)}Erf(\sqrt{v}).
\label{mue2}
\eea

The reader can easily check that eq.(\ref{mue1}) and (\ref{mue2}) are special cases of  our
general expression (\ref{general}). 

In fact the  integral over the proper time $v$ in eq. (\ref{typical}), dropping the L-dependent part in the exponent, is elementary in these two
 cases. For the j=2 case one finds easily, by plugging eq. (\ref{mue2}) into 
eq. (\ref{typical}):
\be
g(r=1,d=3)/L^2={1\over{32\pi}}(12+3\log 3).
\ee

For j=1 one finds:
\be
g(r=1/3,d=3)/L^2={1\over{32\pi}}(4+3\log 3).
\ee

 For interpolating values of $j$ we did the integration numerically 
(see fig. (\ref{fig:gofr})).

\subsection{The zero mode at $j=2$}
We recall eq. ( \ref{typical}):
\begin{eqnarray}
\nonumber g(j,d) & = &{-1\over 2} (Lm_D/4)^{(d-1)}\int{dv\over v}(\pi v)^{{-d+1\over 2}}\big(\sum_n\exp{(-n^2(Lm_D/4)^2/v)}\big)^{(d-1)}\\ \nonumber
       & \times &\big(\exp{-\widetilde E^+_b(j)v}+\Theta(j-1)\exp{-\widetilde E_b^-(j)v}\\ 
       & + & \int_0^{\infty} {dk\over{\pi}}\big({\partial\delta(j,k)\over{\partial k}}-(1-\delta_{j,[j]+1}){\pi\over 2}\delta(k)\big)\exp{-(k^2+4)v}\big). 
\label{typical1}
\end{eqnarray}

As long as the discrete energies are positive (true for $0<j<2$) we can drop the factor $\big(\sum_n\exp{(-n^2(Lm_D)^2/v)}\big)^{(d-1)}$ in eq. (\ref{typical}), because the integral over
$t$ is cut-off by the discrete energy, so every term in the series is bounded by $\exp{(-n^2(Lm_D)^2E_b/4)}$.  That permits us to do the large $L$ limit {\it
before} doing the $v$ integration.

Hence in the infinite volume limit they are all zero, except the term with $n=0$ . And this is the way we compute the tension for $j<2$.

But for $j=2$ we have the positive parity bound state $\tilde E_b^+(j=2)=4-j^2=0$. Then any term in the series with $n^2$ up to $\sim t$ will contribute,
canceling the factor $(\pi v)^{{-d+1\over 2}}$. For large $L$ the result is effectively that obtained in $d=1$ and  we have a logarithmic divergence for $v$ large.

 This is in line with the fact, that in $d=1$  we are computing the quantum corrections to the energy of a classical lump. The uncertainty in the location of the lump at rest is at the origin of the divergence in the rest energy. But the zero energy bound state term (``zero mode'') is needed to give a correct ultra-violet behaviour at $v=0$. We can {\it not}  just drop the zero mode,
 because then Levinson's theorem would predict an u.v. divergence! 

Muenster~\cite{muenster}  introduced a regulator mass, and subtracted in the large volume limit 
the ensuing logarithm. For our purpose we just take as regulator the mass
of the positive parity bound state. So we do the large L limit before the eigentime integration, and calculate the tension  as the limit of the integral for
$j<2$. This integral is trivial to do and gives the same analytic result 
as that of Muenster~\footnote{The actual value quoted in ref.~\cite{muenster}
is different because of finite renormalizations in Muenster's model. In our case no finite renormalizations occur (see I). We thank
Gernot Muenster for telling us his result in terms of the bare variable $m_D$.  }.

For $j>2$ the integration over the proper time $v$ starts to diverge because
the positive parity bound state develops a negative mass.

\section{Predictions}\label{sec:predict}

The basic result is the calculation of $g(r,d=3)$ in eq.(\ref{typical}).
It is shown below in fig. (\ref{fig:gofr}).

\begin{figure}
\begin{center}
\includegraphics{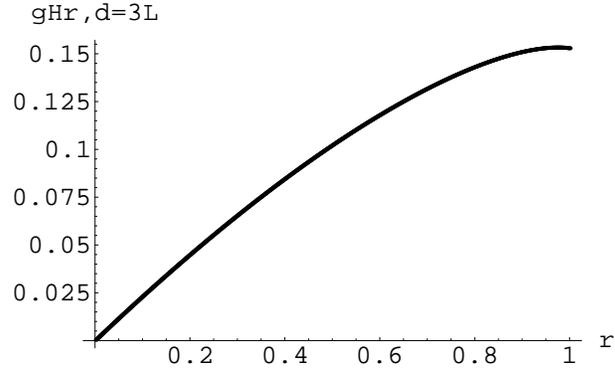}
\caption{The function $g(r,d=3)$ from eq. (\ref{typical}).}
\label{fig:gofr}
\end{center}
\end{figure}

From the numerical values in this figure and eq. (\ref{eq:logdetcu}), relating 
$g(r,d=3)$  to $\rho^{(3)}_k$ we get 
\be
\rho^{(3)}_k= (k^2-1)g(1-r,d=3)+((N-k)^2-1)g(r,d=3)+g(1,d=3).
\label{formuledelatension}
\ee

In the table below we list for the $N\le 8$ our prediction for the cubic correction, and our old results for lowest order and $O(g^2)$.

\begin{table}
\begin{center}
\begin{tabular}{|c|c|c|}
\hline
$N$& $k$ & ${\rho^{(3)}_{k}\over m_d^2}$\\ 
\hline
3& 1& 0.370...\\ 
\hline
4& 1& 0.598...\\ 
 & 2& 0.769...\\ 
\hline
5& 1& 0.831...\\ 
 & 2& 1.190...\\ 
\hline
6& 1& 1.066...\\ 
 & 2& 1.624...\\ 
 & 3& 1.797...\\ 
\hline
7& 1& 1.301...\\ 
 & 2& 2.068...\\ 
 & 3& 2.422...\\ 
\hline
8& 1& 1.538...\\ 
 & 2& 2.519...\\ 
 & 3& 3.063...\\ 
 & 4& 3.236...\\ 
\hline
\end{tabular}
\caption{Cubic contribution to the tension for $N \le 8$} 
\end{center}
\end{table}

\section{Comparison with gradient expansion and large N results}\label{sec:compare}

Our large N results in I were obtained starting from eq. (\ref{eq:logdetcu}):
\bea
\nonumber \rho_k^{(3)}&=&{1\over 2}(k^2-1)\Bigg(\log\det\Big(-\vec\partial^2+m_D^2(1+6(1-r)q(q-1))\Big)-{1\over 2}
\log\det(-\vec\partial^2+m_D^2)\Bigg)\\
 \nonumber &+&k\leftrightarrow N-k\\ 
  &+&{1\over 2}\log\det\Big(-\vec\partial^2+m_D^2(1+6q(q-1))\Big)-{1\over 2}\log\det\Big(-\vec\partial^2+m_D^2\Big).
\label{logdetcu1}
\eea

We then used the gradient expansion: as in ref.~\cite{bhatta} and I  we neglect the z-dependence in the profile $q(z)$. But then the q-dependent Debye mass
in (\ref{logdetcu1}) should be non-negative:
 only those terms on the r.h.s. with $1+6rq(q-1)$ and $1+6(1-r)q(q-1)$ non-negative for all values of $q$ are allowed.  That restricts $r=k/N$ to $1/3\le r\le 2/3$. These terms 
are appearing in the leading order in a large N expansion, where we keep $r$ fixed, instead of $k$ fixed. 

In that limit the cubic correction to the effective potential becomes:
\bea
\nonumber V^{(k)}_{3}&=&{1\over 2}N^2r^2\Bigg(\log\det\Big(-\vec\partial^2+m_D^2(1+6(1-r)q(q-1))\Big)-{1\over 2}
\log\det(-\vec\partial^2+m_D^2)\Bigg)\\ 
 &+&r\leftrightarrow 1-r.
\label{logdetcu1largen}
\eea

 This permits us to do the trace by taking the free particle density of states and doing the integral. That gives us a factor $L^3/(4\pi t)^{d/2}$. Doing the t-integrations gives us a factor $m_D^3\Gamma(-3/2)\Big((1-6rq(q-1))^{3/2}-m_D^3\Big)$. So $V^{(k)}_{3}$ will be of order $m_D^3$.
The total effective action becomes to this order (see ref. I):

\be
S_{eff}^{(k)}(q)=\int_{-\infty}^{\infty} dzK^{(k)}(q)(\partial_zq)^2+V^{(k)}_1+V^{(k)}_2+V^{(k)}_3+...
\label{effaction2}
\ee

\noindent with $V^{(k)}_3$ as in eq. (49) of I.

As in I we minimize the action by varying the profile $q$. The value of the action at the minimum $q_m$ is then the tension $\rho_k$. Doing this leaves us with the result:
\be
\rho^{(3)}_k=\int_{-\infty}^{\infty} dz V^{(k)}_3(q_m(m_Dz))
\label{z}
\ee
\noindent like in I, eq.(51 and (52)). The only difference is the change of variables from z to q. The equation of motion following from the minimization
gives us precisely the relation ${dq\over{dz}}=V_1$ between eq.(\ref{z})) above, and (52) in I. 

Note that the minimum profile only depends on the slow variable $m_Dz$.  As the potential is $O(m_D^3)$ it follows that $\rho^{(3)}_k$ is $O(m_D^2)$.

\begin{figure}
\begin{center}
\includegraphics{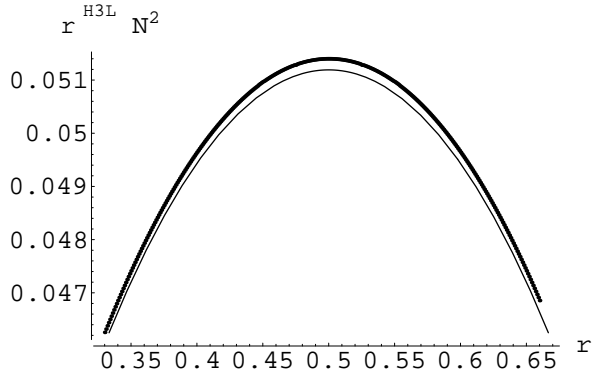}
\caption{Gradient expansion (thin line) versus exact computation (bold line) for large $N$. }
\label{fig:rho2}
\end{center}
\end{figure}

In fig. (\ref{fig:rho2}) we compare the large N result from this paper
with the gradient approximation in the region where it makes sense.

As we can see in eq.(\ref{formuledelatension}), for large $N$ and finite
$r$ $(={k\over N})$, the leading term in the tension is of order $N^2$. As $N$ goes to infinity, $\rho^{(3)} \over N^2$ is a function of $r$, plotted in
fig~\ref{fig:rho2}. So, in this figure, corrections of $O({1\over N^2})$ have
been neglected and only the leading constant term in $N$ of
$\rho^{3} \over N^2$ is plotted.  We see
that the gradient expansion result is in very good agreement with the exact
result in the large N limit (There is about half a percent difference.).

\section{Conclusions}
The loop  can be measured in lattice simulations~\cite{deforluc} and in particular ratio's of different strengths are relatively easily obtained. 
From the table one can easily infer that Casimir scaling of the one and two loop result still holds to a good approximation, even down to the critical temperature.

\begin{figure}
\begin{center}
\includegraphics{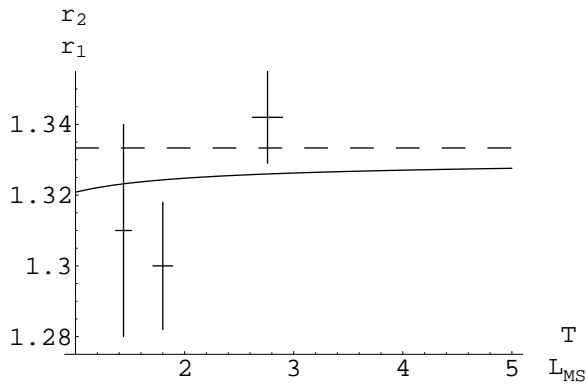}
\caption{$\rho_2 \over \rho_1$ up to $g^3$ as a function of $T\over{\Lambda_{\overline{MS}}}$. Dashed line is
  Casimir scaling, valid for the one loop and two loop prediction. }
\label{fig:rho}
\end{center}
\end{figure}

Before comparing to lattice data a general comment. Our prediction is in terms
of  $T/\Lambda_{\overline{MS}}$, the lattice data in terms of $T/T_c$. Only for N=2 and 3 the ratio $T_c/\Lambda_{\overline{MS}}$ has been determined. So for SU(4) data we take an uncertainty of about 5 $\%$ for that ratio. 


In fig. (\ref{fig:rho}) we compare the one loop, two loop  and cubic order predictions to the lattice  data from de Forcrand et al.~\cite{deforluc} at various temperatures for the SU(4) ratio $\rho_2/\rho_1$.  The horizontal error bars represent our lack of knowledge of the ratio $T_c/\Lambda_{\overline{MS}}$.  

Thus the 't Hooft loop k- ratio is a quantity, that converges well in almost all of the deconfined phase for $N\ge 4$. 

There is long standing evidence from lattice data that convergence is good, once the first coefficient from the magnetic sector is included. Often this coefficient is the dominant contribution for reasonable values of $T\ge 2T_c$, like for the Debye screening mass~\cite{lainekajantie}. As we argued in I, for the 't Hooft loop this contribution will come into play only at $O(g^5)$, or higher. 
For the ratio's these contributions are apperently not very important. 
It may be that for the loops themselves this correction is needed, to have
convergence for $T\ge 2T_c$.

\section*{Acknowledgments}

Discussions  with   Philippe de Forcrand, Bob Jaffe, Mikko Laine, Gernot Muenster, Paul Sorba and Mike Teper were of great value to us.  The hospitality and financial support of the Department of Theoretical Physics of Bielefeld University is gratefully acknowledged by the authors.

\section*{Appendix A. Phase shifts}

In this appendix the phase shifts in the positive and negative parity channels of the Schroedinger equation eq. (\ref{schroe}) are given. The reason for this 
long appendix is  that in previous work~\cite{dashen}~\cite{muenster}  only the solution for $r=1$ was discussed. This is an exceptional case where reflection absent. In this work we needed all values between $0<r\le 1$. Hence reflection is playing a role in the discussion. At the same time we have the anomalous
relation between density of states~\cite{barton}.

In terms of $\zeta=zm_D/2$ eq. (\ref{schroe}) becomes

\begin{equation}
\big(-\partial_{\zeta}^2+6r\tanh^2(\zeta)\big)\psi(\zeta)=({4\over{ m_D^2}}E+6r-4)\psi(\zeta)
\label{morse}
\end{equation}
\noindent and equals q. 12.3.22 of ref.~\cite{morseref}.

It is customary to write $6r=j(j+1)$. As  $j$ grows,  at each integer value of 
j a bound state of parity $(-)^j$ enters. 

Define  the shorthand
\begin{equation} 
\epsilon={4\over{ m_D^2}}E+6r-4. 
\label{eps}
\end{equation}

Then, from \cite{morseref} we have for the n'th discrete eigenvalue, $n<j$:

\begin{equation}
\epsilon_n=6r-\big(\sqrt{6r+{1\over 4}}-(n+{1\over 2})\big)^2=j(j+1)-(j-n)^2.
\label{discreteev}
\end{equation}

It is convenient to use $j=\sqrt{6r+{1\over 4}}-{1\over 2}$.

Using eq.(\ref{eps}) we get 
\be
E_n(j)={m_D^2\over 4}(4-(j-n)^2).
\ee

We now turn to the continuous spectrum of (\ref{morse}).

First a few general remarks about parity invariant scattering channels.
The wave vector $k$ appears in the asymptotic behaviour $exp{(ik\zeta+i\delta_T^{(r)}(k))}$ and  $exp{(-ik\zeta+i\delta_R^{(r)}(k))}$ of the phase shifts of the transmitted ($\delta_T^{(r)}(k)$) and the reflected waves ($\delta_R^{(r)}(k)$). So the wave vector $k$ is related to the momentum $p$ in the phase $exp{(ipz)}$ by $p={m_Dk\over 2}$.

We write the wave function with the transmitted and reflected waves as:
\begin{eqnarray}
\psi^{(l)}(\zeta) & = & \exp{(ik\zeta)}+Rexp{(-ik\zeta+i\delta_R^{(r)}(k))}\mbox{ for}~~ \zeta <<1\\
\psi^{(l)}(\zeta) & = & Texp{(ik\zeta+i\delta_T^{(r)}(k))}\mbox{ for}~~ \zeta >>1.
\label{rlasy}
\end{eqnarray}

The superscript $l$ indicates the wave travels from the left to the right.  $R$ and $T$ are non-negative real numbers, the reflection and transmission coefficients respectively. 

We can define a linearly independent solution with the same wave vector $k$, that travels from the right to the left. This is just the parity transformation $P$, with $P\psi^{l}=\psi^{r}$.
\begin{eqnarray}
\psi^{(r)}(\zeta) & = & \exp{(-ik\zeta)}+Rexp{(ik\zeta+i\delta_R^{(r)}(k))}\mbox{ for}~~ \zeta >>1\\
\psi^{(r)}(\zeta) & = & Texp{(-ik\zeta+i\delta_T^{(r)}(k))}\mbox{ for}~~ \zeta <<1.
\label{rlasy2}
\end{eqnarray}

Alternatively we can define the positive and negative parity continuum states
with (we drop the $r$ and $k$ dependence in the subsequent formulae):
\begin{eqnarray}
\psi^{(+)}(\zeta) & = & \cos(k\zeta+\delta^{(+)})\mbox{ for}~~ \zeta >>1 \\
\psi^{(-)}(\zeta) & = & \sin(k\zeta+\delta^{(-)})\mbox{ for}~~ \zeta >>1.
\label{plusminasy}
\end{eqnarray}
defined to be even (odd) in $\zeta$.

The relation between the two bases is given by:
\begin{equation}
\psi^{(l)}=\exp{i\delta^{(+)}}\psi^{(+)}+i\exp{i\delta^{(-)}}\psi^{(-)}
\end{equation}
\noindent and the parity transform of this equation.

From equations (\ref{plusminasy}) and (\ref{rlasy}) one finds easily
that the reflexion phase shift is related to the transmission phase shift by:
\begin{equation}
\exp{i(\delta_R-\delta_T)}=i\epsilon(\sin(2(\delta^{(+)}-\delta^{(-)})))
\label{difference}
\end{equation}

\noindent and that
\begin{equation}
\exp{i(\delta_R+\delta_T)}=\exp{(i(\delta^{(+)}+\delta^{(-)}))}i\epsilon(\sin(2(\delta^{(+)}-\delta^{(-)})))
\label{sum}
\end{equation}

From these two equations follows that, as long as $\sin(2(\delta^{(+)}-\delta^{(-)}))\neq 0$, we have the identity:

\begin{equation} 
\exp{i(\delta^{(+)}+\delta^{(-)})}=\exp{i\delta_T}.
\label{plusmintrans}
\end{equation} 

Only when the positive parity and negative parity phase shifts do lag behind each other by multiples of ${\pi\over 2}$ this relation might be invalidated.

 This happens typically when the 
potential changes from being flat ($r=0$) to a shallow well ($r>0,r<<1$. The well then produces only the positive parity bound state. This state is formed from a $\cos(k\zeta)$ plane wave with k=0. The parity minus $\sin(k\zeta)$ plane waves stay plane waves, because the well is not yet deep enough to produce the
$tanh(\zeta)$ bound state. But they pick up a shorter wavelength due to the well, so are pulled into the well region.

There is an alternative derivation, simplyfying this condition.

From the definitions of the phase shifts follows:
\begin{equation}
\nonumber 
T=|T|\exp{i\delta_T}={1\over 2}(\exp{i2\delta_+}+\exp{i2\delta_-})\\
R=|R|\exp{i\delta_R}={1\over 2}(\exp{i2\delta_+}-\exp{i2\delta_-})
\end{equation}

Then:
\begin{equation}
T\pm R=\exp{2i\delta_{\pm}}.
\label{trvsdelta}
\end{equation}

That is: as long as $T\neq 0$, we factor out $T$ and find:
\begin{equation}
2\delta_{\pm}=\delta_T+Im\log(1\pm {R\over {T}}).
\label{trvsdeltabis}
\end{equation}

So 
\begin{equation}
2(\delta_{+}+\delta_{-})=2\delta_T+Im\log(1-({R\over T})^2)
\label{trvsdeltabisbis}
\end{equation}

and the last term is zero because 
\begin{equation}
-({R\over T})^2=-{|R|\over{|T|}}(i sign \sin(2(\delta_{+}-\delta_{-}))^2\ge 0.
\end{equation}

The case that $T=0$ ($|R|=1$) is only realized for $j\neq 1,2$ and there {\it only} for $k=0$.  yields for $k=0$:
\begin{equation}
\pm\exp{i\delta_R}=\exp{2i\delta_{\pm}}.
\end{equation}

Hence:
\begin{equation}
\delta_{+}-\delta_{-}=\pm {\pi\over 2}.
\end{equation}
 Here the sign depends on the value of $j$ as will be shown in Appendix B.

\section{Appendix B.Density of states in terms of phase shifts}

We put bc's like in Barton~\cite{barton}:

\begin{equation}
\cos(p_nL+\delta_+)=0,~~\mbox{or}~~ p_nL+\delta_+=\pi(n+1/2)
\end{equation}
for the positive parity channel
and 
\begin{equation}
\sin(p_nL+\delta_-)=0,~~\mbox{ or} ~~ p_nL+\delta_-=\pi n.
\end{equation}
for the negative parity channel.

We start with the positive parity channel. We take $j\neq 1,2$.
We find for the sum over states with potential, compared to the one without
potential:
\begin{equation}
\exp{-E^+_b(j)t}+\int{dk\over {\pi}}\ Big({d\delta_+(k)\over{dk}}-{\pi\over 2}\delta(k)\Big)\exp{(-t{m_D^2\over 4}(k^2+4))}.
\end{equation}

For $t=0$ we find for this quantity $1-{\delta_+(0)\over{\pi}}-1/2=0$ for  $j\neq 1,2$.
\begin{equation}
4E^+_b(j)/m_D^2=4-(j)^2, j\le 2.
\end{equation}

For the negative parity channel we have:

\begin{equation}
\theta(j-1)\exp{-E^-_b(j)t}+\int{dk\over {\pi}}{d\delta_-(k)\over{dk}}(\exp{-t{m_D^2\over 4}(k^2+4)}.
\end{equation}
Here 
\begin{equation}
4E^-_b(j)/m_D^2=4-(j-1)^2, ~1\le j\le 3.
\end{equation}

For the specific potential we have the phase shifts are now given.
For general $j$ we use ref.(~\cite{morseref}):
\begin{eqnarray}
T   & = & \Gamma(j+1-ik)\Gamma(-j-ik)/(\Gamma(1-ik)\Gamma(-ik))\\
{R\over T} & = & \Gamma(1-ik)\Gamma(ik)/(\Gamma(j+1)\Gamma(-j))
\end{eqnarray}

Using Hankels formula,
\begin{equation}
\Gamma(z)\Gamma(1-z)={\pi\over {\sin(\pi z)}},
\label{morse}
\end{equation} 

\noindent to give all arguments in the  Gamma funtions a positive real part, one gets for the transmission and reflection:
\begin{eqnarray}
T    & = &{\Gamma(1+j-ik)\over {\sin(\pi(-j-ik))\Gamma(1+j+ik)}}{\Gamma(1-ik)\sin(\pi (-ik))\over{\Gamma(1+ik)}}\\  
{R\over T} & = &{\pi\over {\sin(\pi ik)}}/({\pi\over {\sin(\pi (-j))}})
\end{eqnarray}

or
\begin{eqnarray}
T & = &{\Gamma(1+j-ik)\Gamma(1+ik)\over{\Gamma(1+j+ik)\Gamma(1-ik)}}\times{\sin(\pi (-ik))\over{\sin(\pi(-j-ik))}}\\ 
{R\over T} & = &{\sin(\pi (-j))\over{\sin(\pi ik)}}
\end{eqnarray}
so for the transmission phase shift:
\begin{eqnarray}
\nonumber \delta_T^{(r)}(k) & = & -Im\log\big(-i\sin{(\pi j)}cotanh{k}+\cos(\pi j)\big)\\
                  & + & Im\big(\log\Gamma(1+j-ik)-\log\Gamma(1+j+ik)+ \log\Gamma(1+ik)-\log\Gamma(1-ik)\big).
\label{morsephase}
\end{eqnarray}

Binet's formula for the logarithm of the Gamma-functions~\cite{whittaker} leaves us with:
\begin{eqnarray}
\nonumber \delta_T^{(r)}(k) & = & -2\int^{\infty}_{0}{du\over u} \sin(ku){(1-\exp{(-ju)})\over{(\exp{(u)}-1)}}\\
\nonumber                  & - &Im\log\big(-i\sin{\pi j}cotanh{k}+\cos{\pi j}\big)\\
                  & = &B(j,k)+f(j,k)
\label{morsephasebinet}
\end{eqnarray}

where the last equality defines the first and second term.

As k becomes very large and positive the ratio of the Gamma functions, using Stirlings formula in eq. (\ref{morse}) (This is why we transformed to arguments with positive real part.) becomes $-2Im\log(-i)$, and the first term  $Im\log\exp{i\pi j}$. So the transmission phase vanishes
as expected for k large and positive. And the same is true for $\delta_{\pm}$
from eq.(\ref{trvsdeltabis}).

The transmission phase is a smooth function of $k$, apart from a jump in $k=0$
in the term  $f(j,k)\equiv -Im\log\big(-i\sin{\pi j}\coth{\pi k}+\cos{\pi j}\big)$.

This function starts for a fixed value of $j$ at large $k$ at the value
$j\pi$:
\begin{equation}
f(j,k)=-Im\log\big(\exp{-i\pi j}+2i\sin(\pi j)\exp{-2k}\big)=\pi j-Im(2i\exp{-i\pi j}\sin(\pi j)\exp{-2k},
\end{equation}
so the asymptotic value is corrected by $-\sin(2\pi j)\exp{-2k}$.

On the other hand, near $k=0$:
\begin{eqnarray}
f(j,k) & = & -Im\log\big(-i\sin{\pi j}\coth{\pi k}+\cos{\pi j}\big)\\
       & + & -Im\log\big(-i\sin{\pi j}({1\over{\pi k}})(1+{1\over 3}(\pi k)^2)(1+icot(\pi j)\pi k(1-{1\over 3}(\pi k)^2)\\
       & = & -Im\log(-i)(1+icot(\pi j)\pi k)
\end{eqnarray}
 the sign of the coefficient $cotg(\pi j)$of $k$ flips at half-integer  values of $j$ with period $\pi$. This renders the half-integer  values of $j$ fixed points when $k\rightarrow 0$.

Hence for $0\le j\le 1$ the value of $f(j,k=0)$ is ${\pi\over 2}$ and $3{\pi\over 2}$ if $1\le j\le 2$, except in the exceptional cases $j=integer$ where there is no dependence of $f(j,k)$ on $k$ and $R=0$. 
It is easy to see that:
\begin{equation}
f(j+1,k)=f(j,k)+\pi
\end{equation}
so are just rigid translates of one another.
This just reflects the growing number of bound states and Levinson's theorem,
that relates the number of bound states to $f(j,k=0)$.

The term $g_{\pm}=Im\log(1\pm R/T)=Im\log\big(1\pm i\sin(\pi j)/\sinh(\pi k)\big)$
so equals $Im\log(\pm i\sin(\pi j))$ at $k=0$. 
The phase shifts with fixed parity are:
\begin{equation}
\delta_{\pm}(j,k)=B(j,k)+f(j,k)+g_{\pm}(j,k)
\end{equation}
Finally, adding up $f$ and $g_{\pm}$ one finds for $0\le j\le 1$ :
\begin{eqnarray}
\delta_+(k=0)& = &{\pi\over 2}\\
\delta_-(k=0)& = &0.
\end{eqnarray}
If $1\le j\le 2$
\begin{eqnarray}
\delta_+(k=0)& = &{\pi\over 2}\\
\delta_-(k=0)& = & \pi.
\end{eqnarray}

This in accord with Barton's anomalous Levinson theorem~\cite{barton} for the positive parity channel in the presence of only one positive parity bound state, and the normal Levinson theorem for the negative parity channel, in the presence of one negative bound state from $\ge 1$ on. 

Let us finally look at the exceptional cases, where j is integer.

Reflection is absent if $j=0, 1, 2$ or $r=0, 1/3, 1$ so $|T|=1$ there.
We have $g_{\pm}=0$ so positve and negative parity phase shifts are identical
and equal to the transmission phase shift $\delta_T$.
 These are precisely the  points where a new bound state enters.
For $r=1/3$, or $j=1$, one finds from eq.(\ref{morsephasebinet}), expanding in the argument $ku$ :
\begin{equation}
\delta_T^{(1/3)}(k)=\pi-2\arctan k
\label{ronethird}
\end{equation}
and for $r=1$ or $j=2$:
\begin{equation}
\delta_T^{(1)}(k)=2\pi-2\arctan k-2\arctan {k\over 2}.
\label{rone}
\end{equation}

As expected from the general reasoning above, for large k the phase shifts vanish. And Levinson's theorem in its normal form is verified: the total number of continuum states, compared to the number of continuum states without the potential,  is
\begin{equation}
\int_{o}^{\infty}{dk\over {\pi}}\delta_T^{(r)}(k)={\delta_T^{(r)}(k=0)-\delta_T^{(r)}(k=-\infty)\over{\pi}}=-n(r)
\end{equation}
\noindent with $n(r)$ the number of bound states, from eqns (\ref{ronethird}) and  (\ref{rone}).

\section*{Appendix C.Relation to error function}

Consider eq. (\ref{typical}):
\begin{eqnarray}
g(j,d) & = &-{1\over 2} (Lm_D/4)^{(d-1)}\int{dv\over v}(\pi v)^{{-d+1\over 2}}\big(\sum_n\exp{(-n^2(Lm_D/4)^2/v)}\big)^{(d-1)}\\ \nonumber
       & \times &\big(\exp{-\widetilde E^+_b(j)v}+\Theta(j-1)\exp{-\widetilde E_b^-(j)v}\\ \nonumber
       & + & \int_0^{\infty} {dk\over{\pi}}\big({\partial\delta(j,k)\over{\partial k}}-(1-\delta_{[j]+1,j}){\pi\over 2}\delta(k)\big)\exp{-(k^2+4)v}\big). 
\label{typical2}
\end{eqnarray}

For $j=1,2$ there is a simple relation to the error function:
\begin{equation}
E(\sqrt{v})={2\over{\sqrt{\pi}}}\int_0^{\sqrt{v}}dx\exp{-x^2}.
\label{errorfunction}
\end{equation}

The derivatives of $E$ and of $I(v)=\int_0^{\infty}{dk\over{\pi}}{2\over{k^2+1}}\exp{-v(k^2+1)}$ with respect to $v$ equal ${1\over{2\sqrt{\pi v}}}\exp{-v}$.
That means:
\begin{equation}
I(v)=c-E(\sqrt{v})
\end{equation}
\noindent with c determined by the behaviour at $v=0$, $c=1$.
For $j=1$ we have $E_b^+(j=1)=3$ and $\exp{(-3v)}I(v)$, hence
\begin{equation}
\exp{(-vE_b^+(j=1))}+\exp{(-3v)}I(v)=\exp{(-3v)}E(\sqrt{v}).
\end{equation}
The r.h.s. has an expansion in $(\sqrt{v})^{2n+1}$, n=0,1,2,...

For j=2  these same terms are there again, except that $\exp{-3t}$ is now furnished by the negative parity bound state with energy $E_b^-(j=2)=4-(2-1)^2=3$. 
 We have also a term $J(v)=\int_0^{\infty}{dk\over{\pi}}{4\over{k^2+4}}\exp{-v(k^2+4)}$, whereas the boundstate with positive parity has become zero:

$$\exp{-vE_b^+(j=2)}=1.$$

 Along the same lines:
\begin{equation}
J(v)=1-E(2\sqrt{v}).
\end{equation}

And so for $j=2$:
\begin{equation}
\exp{(-vE_b^+(j=2))}+\exp{(-vE_b^-(j=2))}+\exp{(-3v)}I(v)+J(v)=E(2(\sqrt{v})+\exp{(-3v)}E(\sqrt{v}).
\end{equation}

Again the r.h.s. is a series in odd powers of $\sqrt{v})$ for small $v$!

For any integer value of j one can easily find  the generalization:
\bea 
\nonumber
& &Tr\exp{-t(\partial_z^2+V(j,z))}-\exp{-t(\partial_z^2+m_D^2)}\\ 
&=&\exp{-4v}\Big(\sum_{n=1}^{j}\exp{(n^2v)}Erf(\sqrt{n^2t})\Big).
\eea
  In this equation the proper time $t$ is related to $v$ by $v=tm_D^2/4$, like before.

\section*{Appendix D.Interpolation}

For interpolating values of j there is the remarkably simple expression (\ref{general})  for  the trace. It is based on the observation that only odd powers of $\sqrt{v}$ survive in the error functions for integer j and that this generalizes to interpolating $j$. This we want to prove below. We write the trace once more below (for $0<j\le 2$):
\begin{eqnarray}
\nonumber & &\big(\exp{-\widetilde E^+_b(j)v}+\Theta(j-1)\exp{-\widetilde E_b^-(j)v}\\ \nonumber
& + & \int_0^{\infty} {dk\over{\pi}}\big({\partial\delta(j,k)\over{\partial k}}-(1-\delta_{[j]+1,j}){\pi\over 2}\delta(k)\big)\exp{-(k^2+4)v}\big)\\ 
&=&{2\over{\sqrt{\pi}}}\int_0^{\infty}du\exp{-u^2}\sinh((j+1)u\sqrt{v}){\sinh(ju\sqrt{v})\over{\sinh(u\sqrt{v})}}+I_0(j,v).
\label{general1}
\end{eqnarray}

Remarkable is that the remainder $I_0(j,v)$ is {\it identically} zero!  Use the expressions for the transmission in eq. (\ref{morsephase}) and (\ref{morsephasebinet}). Integrate the transmission  over $k$. 
This leads for the remainder to:
\bea
\nonumber \exp{(4v)}I_0(j,v)&=&\sum_{n=0}^{[j]}\exp{((j-n)^2v)}-{1\over{\sqrt{\pi}}}\int_{\infty}^{\infty} du\exp{-u^2} cosh((j+1)u\sqrt{v}){\sinh(ju\sqrt{v})\over{\sinh(u\sqrt{v})}}\\ 
&-&(1-\delta_{j,{j}+1}){\pi\over 2}-\int_0^{\infty} dk {sin(2\pi j)\over{sin^2(\pi j)+sinh^2(\pi k)}}\exp{-k^2v}.
\label{identity}
\eea
The last term contains  the part of the transmission phase shift $f(j,k)$ defined in (\ref{morsephasebinet}). It vanishes for integer $j$. So does the one but last term. Obviously only integer powers of $v$ appear in the r.h.s. for any j.

Use now the identity, valid for j integer:
\be
{\sinh(ju\sqrt{v})\over{\sinh(u\sqrt{v})}}=\exp{(-(j-1)u\sqrt{v})}\sum_{n=0}^{j-1}\exp{2nu\sqrt{v}}.
\ee
 
Substitute this in eq. (\ref{identity}). The second term then reduces to the 
sum over bound states in the first term. So for integer $j$ we have indeed $I_0(j,v)=0$. 

For non-integer $j$ the coefficients of $v^l$  can be shown  to vanish for all $l$. This proof is  related to that of the  moment sumrules in ref.\cite{jaffe}.

We start by splitting the phase shift $B(j,k)$ in eq. (\ref{morsephasebinet}) into $B(j,k)=B_s(j,k)+B_a(j,k)$ with:

\be
 B_s(j,k)=\int_0^{\infty}{du\over u} sin(ku)\Bigg({(1-\exp{(-ju)})\over{(1-\exp{(-u)})}}+  {(1-\exp{(ju)})\over{(1-\exp{(u)})}}\Bigg),
\label{split}
\ee
 and for $B_a$ we have the same but with a minus sign in between the two terms on the r.h.s..

After integration over k $B_s$ gives the $cosh$ term in eq. (\ref{identity}) for $I_0(j,v)$, and 
$B_a$ the $sinh$ term in eq.(\ref{general1}). We define $\delta_s(j,k)=B_s(j,k)+f(j,k)$, and 
$\delta_a(j,k)=B_a(j,k)$ so that $\delta(j,k)=\delta_s(j,k)+\delta_a(j,k)$. Note that $\delta_a(j,k=0)=0$.

Then the coefficient $I_0^{(l)}(j)$ of $v^l$ in the expansion of $I_0(j,v)$ reads for $l>0$:
\be
I_0^{(l)}(j)=\sum_{n=0}^{[j]}(j-n)^{2l}+\int{dk\over {\pi}}k^{2l}{\partial\over{\partial k}}\delta_s(j,k).
\label{coeff}
\ee

For $l=0$ we get as right hand side in eq. (\ref{coeff}):
$$n_b-{1\over 2}(1-\delta_{j,[j]+1})+{\delta_s(j,k=0)\over{\pi}},$$
\noindent  which is nothing but Levinson's theorem, eq. ({\ref{levinson}) so vanishes.

For $l\ge 1$ we use two facts:
\begin{itemize}
\item{${\partial\over{\partial k}}\delta_s(j,k)$ has the poles corresponding to the bound states in the upper half plane (For integer j we just proved this.).}
\item{$\delta_s(j,k)$ falls of faster than any power of k.}
\end{itemize} 

Thus one can close the contour in the upper half plane, and the coefficient vanishes through the cancellation of the sum  and the sum of the residues in the integral .

\end{document}